\input jytex.tex   
\typesize=10pt \magnification=1200 \baselineskip=17truept
\hsize=6truein\vsize=8.5truein

\sectionnum=0
\sectionnumstyle{arabic}
\chapternumstyle{blank}
\chapternum=1
\pagenum=0

\def\begintitle{\pagenumstyle{blank}\parindent=0pt\begin{narrow}[0.4in]}
\def\endtitle{\end{narrow}\newpage\pagenumstyle{arabic}}


\def\beginproblem{\vskip 20truept\parindent=0pt\begin{narrow}[10 truept]}
\def\endproblem{\vskip 10truept\end{narrow}}


\def\eql#1{\eqno\eqnlabel{#1}}

\def\ref{\reference}
\def\peq{\puteqn}
\def\pref{\putref}

\def\mgn{\marginnote}
\def\bpr{\begin{problem}}
\def\epr{\end{problem}}


\font\open=msbm10 
 
\def\mbox#1{{\leavevmode\hbox{#1}}}
\def\hspace#1{{\phantom{\mbox#1}}}
\def\oZ{\mbox{\open\char90}}

\def\oN{\mbox{\open\char78}}

\def\al{\alpha}

\def\be{\beta}
\def\ga{\gamma}
\def\de{\delta}

\def\la{\lambda}
\def\La{\Lambda}
\def\om{\omega}

\def\th{\theta}
\def\ze{\zeta}

\def\De{\Delta}

\def\Real{{\rm Re\,}}
\def\Imag{{\rm Im\,}}
\def\zf{$\zeta$-- function}
\def\zfs{$\zeta$-- functions}
\def\hk{heat--kernel}


\def\frac#1/#2{\leavevmode\kern.1em
\raise.5ex\hbox{\the\scriptfont0 #1}\kern-.1em/\kern-.15em
\lower.25ex\hbox{\the\scriptfont0 #2}}
\def\sfrac#1/#2{\leavevmode\kern.1em
\raise.5ex\hbox{\the\scriptscriptfont0 #1}\kern-.1em/\kern-.15em
\lower.25ex\hbox{\the\scriptscriptfont0 #2}}

\def\gtorder{\mathrel{\raise.3ex\hbox{$>$}\mkern-14mu
             \lower0.6ex\hbox{$\sim$}}}
\def\ltorder{\mathrel{\raise.3ex\hbox{$<$}|mkern-14mu
             \lower0.6ex\hbox{\sim$}}}

\def\semidirprod{\rlap{\ss C}\raise1pt\hbox{$\mkern.75mu\times$}}
\def\for{\lower6pt\hbox{$\Big|$}}
\def\fish{\kern-.25em{\phantom{abcde}\over \phantom{abcde}}\kern-.25em}

 

\def\boxit#1{\vbox{\hrule\hbox{\vrule\kern3pt
        \vbox{\kern3pt#1\kern3pt}\kern3pt\vrule}\hrule}}
\def\dalemb#1#2{{\vbox{\hrule height .#2pt
        \hbox{\vrule width.#2pt height#1pt \kern#1pt
                \vrule width.#2pt}
        \hrule height.#2pt}}}


\def\noin{\noindent}


\def\viz{{\it viz.}}

\def\ie{{\it i.e. }}
\def\cf{{\it cf }}

  

  %

\def\nubar{\overline\nu}

\def\pbar{{\overline p}}

\def\sumdash#1{{\mathop{{\sum}'}_{#1}}}
\def\sumdash2#1#2{{\mathop{{\sum}'}_{#1}^{#2}}}

\def\3j#1#2#3#4#5#6{\left\lgroup\matrix{#1&#2&#3\cr#4&#5&#6\cr}
\right\rgroup}

\def\m?{\mgn{?}}


\def\beginexercise{\vskip 20truept\parindent=0pt\begin{narrow}[10 truept]}
\def\endexercise{\vskip 10truept\end{narrow}}

\def\beginremark{\vskip 10truept\parindent=0pt\begin{narrow}[10 truept]}
\def\endremark{\vskip 10truept\end{narrow}}

\def\beginexample{\vskip 10truept\parindent=0pt\begin{narrow}[10 truept]}
\def\endexample{\vskip 10truept\end{narrow}}

\def\bex{\begin{exercise}}
\def\eex{\end{exercise}}
\def\ber{\begin{remark}}
\def\eer{\end{remark}}
\def\bexa{\begin{example}}
\def\eexa{\end{example}}


\def\aop#1#2#3{{\it Ann. Phys.} {\bf {#1}} (19{#2}) #3}

\def\cmp#1#2#3{{\it Comm. Math. Phys.} {\bf {#1}} (19{#2}) #3}
\def\cqg#1#2#3{{\it Class. Quant. Grav.} {\bf {#1}} (19{#2}) #3}

\def\jmp#1#2#3{{\it J. Math. Phys.} {\bf {#1}} (19{#2}) #3}
\def\jpa#1#2#3{{\it J. Phys.} {\bf A{#1}} (19{#2}) #3}

\def\np#1#2#3{{\it Nucl. Phys.} {\bf B{#1}} (19{#2}) #3}
\def\pl#1#2#3{{\it Phys. Lett.} {\bf {#1}} (19{#2}) #3}

\def\prD#1#2#3{{\it Phys. Rev.} {\bf D{#1}} (19{#2}) #3}

\def\cras#1#2#3{{\it Comptes Rend. Acad. Sci. (Paris)} {\bf{#1}} (#2) #3}

\def\invm#1#2#3{{\it Invent. Math.} {\bf {#1}} (19{#2}) #3}
\def\jdg#1#2#3{{\it J. Diff. Geom.} {\bf {#1}} (19{#2}) #3}

\def\tams#1#2#3{{\it Trans. Am. Math. Soc.} {\bf {#1}} (19{#2}) #3}

\begin{title}
\vglue 1truein
\vskip15truept
\centertext {\Bigfonts \bf Effective actions on squashed lens
spaces} \vskip3truept \centertext{\Bigfonts \bf } \vskip 20truept
\centertext{Marcelo De
Francia\footnote{dfrancia@a13.ph.man.ac.uk},\quad Klaus
Kirsten\footnote{klaus@a13.ph.man.ac.uk}\quad and\quad J.S.Dowker
\footnote{dowker@a13.ph.man.ac.uk}} \vskip 7truept
\centertext{\it Department of Theoretical Physics,\\
The University of Manchester, Manchester, England} \vskip10truept
\vskip10truept \vskip7truept \vskip 20truept \centertext
{Abstract} \vskip10truept
\begin{narrow}As a technical exercise with possible relevance to
the holographic principle and string theory, the effective
actions (functional determinants) for scalars and spinors on the
squashed three-sphere identified under the action of a cyclic
group, $\oZ_m$, are determined. Especially in the extreme oblate
squashing limit, which has a thermodynamic interpretation, the
high temperature behaviour is found as a function of $m$. Although
the intermediate details for odd and even $m$ are different, the
final answers are the same. A thermodynamic interpretation for
spinors is possible only for twisted periodicity conditions and
$m$ even.
\end{narrow}
\vskip 5truept
\vskip 60truept
\vfil
\end{title}
\pagenum=0
\section{\bf Introduction}
Quantum field theory on the squashed three--sphere is of current
interest in connection with the holographic principle but has an
intrinsic technical value and this is the aspect that we mainly
wish to promote here. The present work amends and extends earlier
work of the last named author. In particular, attention is drawn
to some errors of calculation in [\pref{Dow2}] which seriously
affect the conclusions there. Especially, the overall sign of the
spinor free energy is reversed, making a high temperature
thermodynamic interpretation impossible. This could have been
anticipated on the grounds that on the simply--connected squashed
three-sphere, $\widetilde{\rm S}^3$, the spinor field will be
single-valued, and hence periodic around the fibering
$\psi$-circle while a thermal interpretation would require
anti-periodicity. This ties in with the topological fact that the
bulk space-time is not spin. In order to define a conventional
spin structure it is sufficient to identify the bulk spacetime
for example by quotienting the $\psi$-circle by the cyclic group
$\oZ_m$ with $m$ even. We write this as $\widetilde{\rm
S}^3/\oZ_m$ and refer to it as a squashed lens space. Once the
space is multiply connected there is the allowed possibility of
twisted, and, in particular, of anti-periodic fields so that a
thermal interpretation for spinors becomes possible. On the
squashed lens space this can occur only for even $m$, which again
fits in nicely with the absence of spinor structures in the bulk
for odd $m$. In fact we find that odd $m$ is technically more
awkward than even $m$, although the final answers are identical.

In this paper we shall be particularly concerned to exhibit the
explicit high-temperature behaviours of the free energies in the
thermodynamic interpretation and their dependence on the
identification parameter, $m$.

\section{\bf Fields on the squashed lens space}
In this section we concentrate on the mode description, and
thence the \zfs, $\ze(s)$, for scalar and spinor fields on squashed
lens spaces but, first, in order to make this paper relatively
self contained, the geometry of the squashed three--sphere is
briefly outlined, see [\pref{Dow2,Dow1}] and references there.
The local geometry of $\widetilde{\rm S}^3$ is determined by the
metric $$ ds^2=(d\th^2+\sin^2\th d\phi^2)+l_3^2(d\psi+\cos\th
d\phi)^2. $$ Oblate deformations correspond to $l_3<1$ and
prolate to $l_3>1$. When the squashing parameter $l_3$ equals
one, the metric is that of a round three--sphere, ${\rm S}^3$, of
radius 2, expressed in Euler angles.

In the unidentified case, the $\psi$-circle has a circumference
of $4\pi l_3$. For the lens space, with $m$ identifications
around this circle, the periodicity is $4\pi l_3/m$ and, in the
thermodynamical interpretation, where $\psi$ plays the role of a
Euclidean time (roughly speaking), this periodicity translates to
an inverse temperature $\be=4\pi l_3/m$. We therefore {\it
define} a free energy $\Phi(\be)$ to equal $\mp \ze'(0)/(2\be)$, $-$
for scalars and $+$ for spinors, and shall seek to compare this, roughly, 
with a free energy on a 2-sphere to which the metric tends as $\be$ becomes
small.

\subsection{\it Scalar field}
We recall that, up to normalisation, the scalar modes on
$\widetilde{\rm S}^3$ are the ${\rm SU(2)}$ representation
matrices

$$\eqalign{ {\cal D}^{(L)}_{NM}(\phi,\th,\psi)&=e^{iN\phi}\,
P^{(L)}_{NM}(\th)\,e^{iM\psi},\cr \big(L=0,1/2,1,&\ldots,\quad
-L\le N,M\le L\big)\cr} \eql{modes}$$ with  eigenvalues
$$
\la_{LM}={(2L+1)^2\over4}-l_3^{-2} \delta^2
M^2 \eql{eigenv} $$ for the
operator $-\De+1/4$, where
$\delta^2 = l_3^2 -1$ has been introduced. The degeneracy  is $n=2L+1$.

Working with a second order differential propagation operator for
the scalar field means that there is considerable latitude in its
choice, in the sense that various covariant quantities can be
added more or less naturally. The most usual addition is a term
proportional to the scalar curvature, $R$. The choice made here is
the one used in [\pref{Dow2}], the sole reason being the relative
simplicity of the eigenvalues. Other selections were briefly
considered in [\pref{Dow2}]. If one were seriously concerned about
conformal invariance in three dimensions then the relevant
operator would be $-\De+R/8$ for which the eigenvalues are $$
\la^{\rm C}_{LM}=\la_{LM}-{l_3^2\over16}. \eql{confeig}$$
Results will be presented for both sets of eigenvalues.

The lens space factoring, $\widetilde{\rm S}^3/\oZ_m$, is
implemented by the identification of $\psi$ and $\psi+4\pi h/m$,
$h=1,2,\ldots,m-1$, which entails, via periodicity, the
restriction $$ {2M\over m}=k,\quad k\in\oZ. \eql{restr}$$

A generalisation is afforded by the field pseudo-periodicity $$
\psi(q\ga)=a(\ga)\psi(q),\quad q=(\phi,\th,\psi),\quad \ga\in
\oZ_m \eql{pperiod}$$ where $a(\ga)$ is a unitary one-dimensional
representation of $\oZ_m$, $$ a(\ga_0^h)=e^{2\pi ihr/m},\quad
h=0,1,\ldots m-1, $$ labelled by $r=0,1,\ldots,m-1$. The action
of the generator, $\ga_0$, of $\oZ_m$ on $q$ is
$q\ga_0=(\phi,\th,\psi+4\pi/m)$. Equation (\peq{pperiod}) leads to
the twisting restriction $$ {2M-r\over m} = k \eql{twrestr}$$
instead of (\peq{restr}).

The simplest example is $m=2$. When $r=0$, $M$, and therefore
$L$, is integral giving periodic fields while when $r=1$, $L$ is
half-odd integral giving anti-periodic, or `twisted', fields. In
the present context we might call the former `thermal' and the
latter `anti-thermal' and we shall restrict the discussion in
this paper to these cases. Anti-thermal bosons mimic thermal
fermions.

In the general case when $m$ is even these two are the only
possibilities for real fields and correspond to $r=0$ and
$r=m/2$. When $m$ is odd, there are no real twisted fields and,
furthermore, the corresponding bulk space-time is not spin. For
this reason we separate even and odd $m$ and begin with the even
and periodic ($r=0$) case. \vskip 5truept \noin{\it The Mode
Lattice}

Rather than construct projections we deal with the consequences
of the restriction (\peq{restr}) directly. It is convenient to
represent the modes in the {\it unidentified} case on a square
lattice and then impose any identification restrictions
geometrically. The overall restriction $|2M|<2L$ is implemented by
confining the lattice to the positive quadrant, choosing the
horizontal axis, $x$, to equal $L-M$ and the vertical one, $y$,
to be $L+M$. Thus the {\it diagonal axis} is $x+y=2L=n-1$ and the
{\it anti-diagonal} one is $y-x=2M$ and has a finite range.
Unidentified modes correspond to {\it every} integer lattice
point in the positive $x,y$ quadrant, including the edges.

The condition (\peq{restr}) is then allowed for by picking only
those square lattice points that coincide with the intersections
of the line $n=$ constant (for each $n=1,2,\ldots$ in turn) with
the $45^\circ$ lines, $2M=km,\,\forall k\in\oZ$. This is best
done for a few $m$ and the general rule inferred. In this way we
come to the residue class mod $m$ decomposition of the odd number
$n=2L+1$ , $$ n=pm+(2\nu+1),\quad p=0,1,\ldots,\infty,\quad
\nu=0,1,\ldots,m/2-1\,, \eql{rescl}$$ which encompasses all
(positive) odd numbers.

It is then easy to see that the limits on $M$ imply that, for a
given $p$, $|k|\le p$, [\pref{Unwin}]. In order to maintain
contact with the development of our previous work, it is better
to write this as $0\le u\le 2p$, where $u=p-k$ so that the scalar
\zf\ reads $$ \ze^{(e)}(s)=m\bigg({2l_3\over
m}\bigg)^{\!2s}\sum_{\nu=0}^{m/2-1}
\sum_{p=0}^\infty\sum_{u=0}^{2p}
{p+\nubar/m\over\big[(p+\nubar/m)^2+\de^2(u+\nubar/m)
(2p-u+\nubar/m)\big]^{s}} \eql{zeta}$$ where we have set
$\nubar=2\nu+1$ for brevity.

Equivalent to these label relations is the residue class
decomposition, $L-M=u(m/2)+\nu$ where $u$ plays the role of $q$
in our earlier computations, [\pref{Dow2,Dow1}].

The case of {\it odd} $m$ seems to present some peculiar
technical difficulties. \mgn{Give odd m?} Equation (\peq{restr})
means that $M$, and hence $L$, can be integer or half odd-integer
so that $n$ can be both even and odd but with some even
omissions. Again, with reference to the mode lattice, decompose
$n=2L+1,$ $$ n=pm+\nubar, \quad p=0,1,\ldots,\infty,\quad
\nubar=2\nu+1, \quad \nu=0,1,\ldots,m-1. \eql{rclodd} $$ Note
that $n$ runs over all the integers {\it except for} the evens,
$2,4,\ldots,m-1$.

The same limits apply as before and defining $q=p/2-M/m$ we  have
$0\le q\le p$. Equivalently, $L-M=qm+\nu$, which, as a check,
reduces to $q$ when $m=1$.

The \zf\ is $$ \ze^{(o)}(s)=m\bigg({2l_3\over
m}\bigg)^{\!2s}\sum_{\nu=0}^{m-1} \sum_{p=0}^\infty\sum_{q=0}^p
{p+\nubar/m\over\big[(p+\nubar/m)^2+4\de^2(q+\nubar/2m)
(p-q+\nubar/2m)\big]^{s}} \eql{zetao}$$ and the calculation is
essentially identical, in structure at least, to the unidentified
case. When $m=1$ we just set $n=p+1$ to regain the scalar
expressions in our earlier paper.
\subsection{\it The Dirac field}
\noin Before analysing the \zfs\ (\peq{zeta}),
(\peq{zetao}), we turn to the spin-half field.

Apart from a possible mass term there is little one can add to the
Dirac operator, in a natural way. To recall, the unidentified
eigenvalues of the Weyl  operator are $$
\om_{\pm}={1\over2}\bigg[{1\over2}l_3\pm[(2L+1)^2
-4M^2l_3^{-2}\de^2]^{1/2}\bigg]
$$ with $-(L\pm1/2)\le M\le L\pm1/2$ and $L\ge0$ for $\om_+$,
$L\ge1/2$ for $\om_-$.

Some relevant analysis in the identified case has been performed
by Gibbons, Pope and R\"omer, [\pref{GPR}] and Pope,
[\pref{Pope}] in connection with boundary terms in the index
theorem. However in these calculations only the two `extreme'
modes were relevant. (See also [\pref{GandH}].) The more recent
mathematical investigations by B\"ar, [\pref{Bar1,Bar2}], detail the spectrum 
and
also that on the squashed odd sphere, S$^{2n+1}$, which is referred to as the
Berger sphere.

On $\widetilde S^3$ the same restriction holds as in the scalar
case, that is, for periodic spinors $$ 2M=0\,\, {\rm mod}\,\, m
\eql{restrs}$$ and the calculation can proceed, more or less, as
before.

There are now positive and negative modes. For the former, the
orbital quantum number $L$ satisfies $L\ge0$ while for the latter
we have $L\ge1/2$ and the corresponding ranges for the projection
number $M$ are $|M|\le L+1/2$ and $|M|\le L-1/2$ respectively. We
consider the effect of the restriction (\peq{restrs}) on the two
sets of modes separately and again begin with even $m$ when we
can easily see that $n=2L+1$ must also be even.

For the positive modes we decompose $n$ as $$ n=pm+2\nu,\quad
p=0,1,\ldots,\infty,\quad \nu=0,1,\ldots,m/2-1\,. \eql{decomp1}$$

Note that this would lead to the inclusion of the value $n=0$
which is not within the mode labelling scheme. However the
degeneracy factor of $n$ allows an extension to this value. Note
also that all even numbers occur.

Referring to the mode lattice, for the positive modes the
diagonal label is $n$, which takes the value $0$ at the origin.
The `antidiagonal' label is $2M$. It is then easy to see that
(\peq{restrs}) is equivalent to $|2M/m|\le p$. For the negative
modes the diagonal label starts at the origin with the value
$n=2$. In this case we use the decomposition $$ n=pm+2\nu',\quad
p=0,1,\ldots,\infty,\quad \nu'=1,2,\ldots,m/2\,, \eql{decomp2}$$
and, defining $u=p-2M/m$, the relevant summations read
$$
\ze^{(e)}_\pm(s)=\bigg({2l_3\over
m}\bigg)^{\!2s}\!\!\sum_{\nu=\nu_\pm}^{m_\pm}
\sum_{p=0}^\infty\sum_{u=0}^{2p}
{mp+2\nu\over\big[[(p+{2\nu\over m})^2+\de^2(u+{2\nu\over m})
(2p-u+{2\nu\over m})]^{1/2}\pm
{l^2_3\over2m} \big]^{2s}} \eql{zetapl}$$
where $m_+=m/2-1$, $m_-=m/2$, $\nu_+=0$ and $\nu_-=1$, 
($+$ referring to the positive modes and $-$ to the negative).

After expanding in $l^2_3/(2m)$, in order to obtain some
cancellation between the positive and negative modes, the
differing summations over the roots of unity have to be adjusted.
In particular in $\ze_+$ we add the term $\nu=m/2$ and subtract
the term $\nu=0$ in order to convert the summation labels into
those for $\ze_-$.

After some algebra, combining $\ze_+$ and $\ze_-$, it is then
seen, expanding in $l^2_3/(2m)$, that the calculation will devolve
upon evaluating, and continuing, the expression $$
f^{(e)}_m(s)=\sum_{\nu=1}^{m/2}\sum_{p=0}^\infty\sum_{u=0}^{2p}
{mp+2\nu\over[(p+2\nu/m)^2+\de^2(u+2\nu/m)(2p-u+2\nu/m)]^s}
\eql{effeven}$$ to be compared with (\peq{zeta}). The precise
relation for $\ze^{(e)}(s)=\ze^{(e)}_+(s)+\ze^{(e)}_-(s)$ is
$$\eqalign{
\ze^{(e)}(s)=2\bigg({2l_3\over m}\bigg)^{2s}\bigg(
m\ze_H(2s-1,l^2_3/2m)-&{l^2_3\over2}\ze_H(2s,l^2_3/2m)+f^{(e)}_m(s)
\cr
&+{l_3^4\over4m^2}s(2s+1)f^{(e)}_m(s+1)+\ldots\bigg)\,,\cr}
\eql{zetaspe}$$
the first two terms coming from the aforementioned adjustment.

For odd $m$  we again consider the positive and negative modes
separately. For the former, reference to the lattice shows that
with $n=mp+2\nu$, $\nu=0,1,\ldots,m-1$ we have $|M/m|\le p/2$,
while for the latter $n=mp+2\nu$ with $\nu=1,2,\ldots,m$ and the
same condition on $M$. Hence, defining $q=p/2-M/m$ we have $0\le
q\le p$ (as in the scalar case) and the \zfs\ are $$
\ze^{(o)}_\pm(s)=\bigg({2l_3\over
m}\bigg)^{\!2s}\sum_{\nu=\nu_\pm}^{m_\pm}
\sum_{p=0}^\infty\sum_{q=0}^p
{mp+2\nu\over\big[[(p+{2\nu\over m})^2+4\de^2(q+{\nu\over m})
(p-q+{\nu\over m})]^{1/2}\pm {l_3^2\over2m}\big]^{s}} \eql{zoddpl}$$
where now $m_+=m-1$ and $m_-=m$.

Similar manipulations as before produce the same expansion,
$$\eqalign{
\ze^{(o)}(s)=2\bigg({2l_3\over m}\bigg)^{2s}\bigg(
m\ze_H(2s-1,l^2_3/2m)&-{l^2_3\over2}\ze_H(2s,l^2_3/2m)
+f^{(o)}_m(s)\cr
&+{l_3^4\over4m^2}s(2s+1)f^{(o)}_m(s+1)+\ldots\bigg)\,,\cr}
\eql{zetaspo}$$
 where $f^{(o)}_m(s)$ is $$
f^{(o)}_m(s)=\sum_{\nu=1}^m\sum_{p=0}^\infty\sum_{q=0}^p
{mp+2\nu\over[(p+2\nu/m)^2+4\de^2(q+\nu/m)(p-q+\nu/m)]^s}\,.
\eql{effodd}$$

\subsection{\it Conformal scalars}

Conformal scalars are similar to spinors in that an expansion in
$l_3^2$ is required. The form of the eigenvalues (\peq{confeig})
shows that the only difference from the small $l_3$ series of the
effective action will be in the $l_3^2$ term. Denoting by
$\zeta^C (s)$ the $\zeta$-function associated with the spectrum
(\peq{confeig}), the relation
between the \zfs\ is $$ \ze^{\rm C}(s)= \ze(s) +{1\over16} s
l_3^2 \ze(s+1) + O(l_3^4) $$ so that the effective actions are
connected by $$ {\ze^{\rm C}}'(0)= \ze^\prime (0) + {1\over16}
l_3^2 \ze(1) + O(l_3^4).$$ \mgn{MORE. ask Marcelo}
Note that $\zeta (1)$ is finite because we are dealing with a manifold
without boundary so that ${\zeta ^{\rm C}}' (0) $ is well defined.
\subsection{\it Twisted spinors}

So far, for both scalars and spinors, only periodic fields have
been analysed. As discussed in the introduction, we are
interested in the case of anti-periodic spin-half fields. That
is, those fields that pick up a factor of $-1$ on going once
round the $\psi$--circle on $\widetilde{\rm S}^3/\oZ_m$. As
outlined in section 2 this corresponds to the choice of $r=m/2$,
with $m$ even, for the integer, $r$, that determines the twisting.

The general condition on the modes leads to the restriction
$2M=km+r$ or, in the anti-periodic case $$
2M=(2k+1){m\over2},\quad k\in\oZ $$ a simple consequence of which
is that $M$ can never equal zero.

A further brief subdivision of the order $m$ is required. Even
$m$'s are either such that $m$ is divisible by 4 (\ie $m/2$ is
even) or such that $m-2$ is divisible by 4 (\ie $m/2$ is even).
For the first case $2M$ is odd while for the latter $2M$ is even.
The two cases are initially treated separately but reference to
the mode lattice soon produces the same mode counting. This shows
that for the positive modes
$$ n=(p+1/2)m+2\nu,\quad
p=0,1,\ldots\infty,\quad\nu=0,1,\ldots,m/2-1\,. $$ The negative
modes have an $n$ shifted by 2, $$ n=(p+1/2)m+2\nu,\quad
p=0,1,\ldots\infty,\quad\nu=1,2,\ldots,m/2. $$ We note that when
$m/2$ is odd (even), $n$ is odd (even).

The restriction on $M$ is $|2M/m|\le p+1/2$ in both cases so
that, defining $u=p+1/2-2M/m$, one has $0\le u\le 2p+1$ and the
\zfs\ are
$$ \ze^{(ap)}_\pm(s)=\bigg({2l_3\over
m}\bigg)^{2s}\!\!\sum_{\nu=\nu_\pm}^{m_\pm}
\sum_{p=0}^\infty\sum_{u=0}^\pbar
{m\pbar/2+2\nu\over\big[[({\pbar\over2}+{2\nu\over m})^2
+\de^2(u+{2\nu\over m})(\pbar-u+{2\nu\over m})]^{1/2}\pm {l^2_3\over2m} 
\big]^{2s}}
\eql{zpltw2}$$ where attention has been transferred to the odd
integer $\overline p=2p+1$, and $m_\pm$ are as in (\peq{zetapl}).

The expansion function is $$
f^{(ap)}(s,m)=\sum_{\nu=1}^{m/2}\sum_{\pbar=1,3,\ldots}^\infty\sum_{u=0}^\pbar
{m\pbar/2+2\nu\over\big[(\pbar/2+2\nu/m)^2+\de^2(u+2\nu/m)(\pbar
-u+2\nu/m)\big]^s}\,. \eql{effap}$$

Looking at the $\pbar$--sum, we can use the useful identity that `sum over
odds $=$ sum over all $-$ sum over evens'. Firstly, thinking of $\pbar$ as
`all', we see that (\peq{zpltw2}) is identical to the {\it odd} $m$
expression (\peq{zoddpl}) with, {\it formally}, $m\to m/2$. Secondly,
taking $p$ as `evens', (\peq{zpltw2}) is identical to the {\it even} $m$
expression (\peq{zetapl}) with $m\to m$ and so there is no need to do any
real further algebra when $m/2$ is odd, the relation being,
$$
\ze^{(ap)}(s,m)=\ze^{(o)}(s,m/2)-\ze^{(e)}(s,m),\quad
(m/2\quad{\rm odd})\,, \eql{zetaap}$$ or, in terms of the $f$
function, $$
f^{(ap)}(s,m)=2^{2s}f^{(o)}(s,m/2)-f^{(e)}(s,m),\quad
(m/2\quad{\rm odd}) \, , \eql{effap2}$$
with obvious notation. A similar conclusion holds for even $m/2$, after
making some formal identifications.
As we will see later, twisted spinors allow for a thermal interpretation.
\begin{ignore}
\section{\bf Thermal interpretation }

We now look at the thermal interpretation of
(\peq{f3dash})\mgn{CHANGE.
 KK and Marcelo}
and (\peq{f3dasho}) and thence of ${\ze^{(ap)}}'(0,m)$ via the
combination $$
{f^{(3\,a\!p)}_m}'(0)={f^{(3\,o)}_{m/2}}'(0)-{f^{(3\,e)}_m}'(0)\,.
\eql{combi}$$ (using the absence of a conformal anomaly).

The relevant expressions can be found in [\pref{Dow2}]. We have
already made some connections in equns (\peq{f3dashe2}) and
(\peq{f3dasho3}). In the high-temperature limit, the finite sum in
(\peq{f3dasho3}) yields higher power contributions and therefore
can ignored if we are interested in only the leading behaviour.
Because of the $m/2$ in (\peq{combi}) we have to evaluate
(\peq{f3dasho3}) at $2\be'$. Hence, when $m/2$ is odd, $$\eqalign{
{f^{(3\,a\!p)}}'(0,\be')&=W_{\rm sp}^{(3)}(2\be') -W_{\rm
sp}^{(3)}(2\be')\bigg|_{\rm even}\cr &=W_{\rm
sp}^{(3)}(2\be')\bigg|_{\rm odd}\,,\cr} \eql{combi2}$$ which is
the answer expected since this is the result for the special case
$m=2$ and extends to any even $m$, with $m/2$ odd, when $\be'$ is
introduced.

The high-temperature behaviour can be extracted from the
expressions in [\pref{Dow2}] where the cases of odd and even $n$
are usefully separated. Referring to section 5 in [\pref{Dow2}]
we have that $$ W_{\rm sp}^{(3)}(2\be')\bigg|_{\rm
even}=2\be'\widetilde F'_{\rm sc}(\be') $$ in terms of the free
energy of `thermally twisted' {\it scalars} on the two-sphere.
{\it Defining} a free energy generally by $$
\Phi(\be)={1\over\be}W(\be') $$ we find that on the squashed lens
space, $\widetilde{\rm S}^3/\oZ_m$ ($m$ even), the free energy of
anti-periodic spinors has the high-temperature limit $$
\Phi(\be)\sim -{3\ze_R(3)\over
\be^3}+{1\over3\be}\log2-{\be\over240}-
{m^2\be\over128\pi^2}\log2+O(\be^3) \eql{frentwsp}$$ which, as
anticipated, has a legitimate thermodynamic sign. Note the absence
of  $\log\be/\be$ and $\ze_R'(-1)/\be$ terms.

\end{ignore}
\section{\bf Continuation of the \zfs.}

We have so far mainly concentrated on setting up the mode
properties in the identified case. The previous papers, [\pref{Dow2,Dow1}],
supply most of the techniques needed for the continuation of the
resulting \zfs. The method for this, which is not unique,
involves a Plana summation applied to the inner summation
followed, if necessary, by a Watson-Sommerfeld procedure applied
to the $p$--sum. A particular arrangements of cuts in the
introduced complex planes, resulting from the vanishing of
denominators, allows the extraction of factors of $sin\pi s$ and
hence a practical evaluation of $\ze'(0)$. Here we exhibit some
details just in the case of the scalar field for $m$ even. The spinor
case presents only a little extra algebra and, as already
mentioned, while odd $m$ is rather more awkward in the
intermediate calculation it finally yields the same $m$
dependence.

As in Ref.~[\pref{Dow2}], we seek to perform the $u$ sum using the
Plana method. Denote, therefore, the last sum in (\peq{zeta}) by
$I$ and, this time, choose the relevant strip of the $u$-plane to
be $-{\nubar \over{m}} \leq \Real u \leq 2 p + {\nubar
\over{m}}$. In the oblate case, $l_3<1$, there are three parts to
the summation formula, $I=I_1 + I_2 + I_3$. They read explicitly

$$I_1= \int_{-\nubar /m}^{2p+\nubar /m}
{{(p+\nubar/m)du}\over{[(p+\nubar/m)^2+\delta^2(u+\nubar/m)
(2p-u+\nubar/m)]^s}},\eql{i1} $$
$$I_2=-4\, \Imag \int_0^{\infty} {{p + \nubar/m}\over{[(p + i
B^\prime)^2 - C^2]^s}} {{dx}\over{1-e^{2 \pi x -2 i \pi
\nubar/m}}},\eql{i2}$$
$$I_3= \int_{\rm C_U} {{du}\over{1-e^{-2 \pi i u} }}
{{p+\nubar/m}\over{[(p+\nubar/m)^2 +\delta^2
(u+\nubar/m)(2p-u+\nubar/m)]^s}}$$ $$-\,\int_{\rm C_L}
{{du}\over{1-e^{2 \pi i u} }} {{p+\nubar/m}\over{[(p+\nubar/m)^2
+\delta^2 (u+\nubar/m)(2p-u+\nubar/m)]^s}}\,. \eql{i3}$$

The first is the integral approximation. In the second we use
$$ B^\prime = -x \delta^2 - i {\nubar \over{m}} = -x\left( l_3^2
-1 \right) - i {\nubar \over{m}} = B- i {\nubar \over{m}},\quad
C^2 = x^2 (1 - l_3^2) l_3^2, $$ 
and the third term comes from
a clockwise contour surrounding the extra cuts that appear
in the $u$-complex plane when $l_3<1$. The three contributions are
dealt with one by one.

Changing the integration variable in (\peq{i1}) we immediately
obtain the first contribution to the $\zeta$-function,
$$\zeta_{e,1} (s)=2m \left({{2 l_3}\over{m}}\right)^{2s}
\left[\sum_{\nu=0}^{m/2-1} \sum_{p=0}^{\infty}
\left(p+{{\nubar}\over{m}}\right)^{-2s+2} \right] \int_0^1
{{dx}\over{(1+4 \delta^2 x (1-x))^s}}. \eql{zeta1} $$ The sum
in square brackets can be converted to the original $n$-mode sum
(see eqn. (2.7)), running, in this case, over odd $n$-values, to
give $$\zeta_{e,1} (s)={8\over{m}} l_3^{2s}
\zeta_H(2s-2,1/2)\int_0^1 {{dx}\over{(1+4 \delta^2 x
(1-x))^s}}.\eql{zeta1r}$$

Three conclusions may be derived straightforwardly from
expression (\peq{zeta1r}). In the first place, its contribution
to the conformal anomaly, $\zeta_{e,1} (0)$, vanishes. Secondly,
its value at $s=1$, needed in the evaluation of the conformal
field effective action, is $O(l_3^2)$ and is negligible in our
present calculation. Thirdly, the Weyl pole at $s=3/2$ correctly
has a volume factor of $1/m$.

Taking the derivative explicitly, we obtain the first
contribution to the effective action, $$\zeta_{e,1}^\prime
(0)={{3}\over{m \pi^2}}\zeta_R (3). \eql{z1p}$$

The second term in the effective action comes from (\peq{i2}). We
proceed to its calculation again using the Plana technique, in
this case to effect the $p$-sum.

Using the integral representation and changing variables
$$\sum_{p=0}^{\infty} {{p+\nubar/m}\over{[(p+i B^\prime)^2 -
C^2]^s}}={1\over{2i}} \int_{\rm L_D} d\zeta {{\zeta -i
B}\over{(\zeta^2- C^2)^s}} \cot\big(\pi(\zeta -i B^\prime
)\big)\eql{spc}$$
where ${\rm L_D}$ runs anticlockwise around the poles at $\zeta=p+i
B^\prime, p\in \oN_0,$ of the last factor.

The integration path can be deformed towards the imaginary axis and
two contributions to $\zeta_e (s)$ appear \viz\  the integral
over the vertical axis,
$$\zeta_{e,(2,1)} (s)=2 m \left( {{2
l_3}\over{m}} \right)^{2s} \Imag \sum_{\nu=0}^{m/2-1}\int_0^\infty
{{dx}\over{1-e^{2 \pi x-2i \pi \nubar/m}}}\times
\hspace{***********}$$
$$\hspace{**********}\left\{e^{-i \pi
s} \int_0^\infty dy {{y- B}\over{(y^2 + C^2)^s}} \coth
\big(\pi(y - B^\prime)\big)\right.+ \eql{zeta21}$$
$$\hspace{**************}\left. e^{i \pi s} \int_0^\infty dy \, {{y+ B}
\over{(y^2 +C^2)^s}}\coth \big(\pi(y + B^\prime)\big)\right\}
$$
and the integration around the cut on the real axis
between $-C$ and $C$,
$$\zeta_{e,(2,2)} (s)= 4 m \sin (\pi s)
\left( {{2 l_3}\over{m}}\right)^{2s} \Imag \sum_{\nu=0}^{m/2-1}
\int_0^{\infty} {{dx}\over{1-e^{2\pi x-2 \pi i \nubar/m}}}\times
\hspace{********}$$
$$\hspace{************}\int_0^{C} dy \, {{(y-i B)}\over{(C^2 - y^2)^s}}
\cot \big(\pi( y - i B^\prime)\big).
\eql{zeta22}$$

We begin by studying $\zeta_{e,(2,1)}(s)$.
First we make the split, $\coth (\pi x) = 1+2/[\exp (2\pi x) -1]$. Due
to the $\nu$-summation several contributions disappear
leading to the expression
$$\zeta_{e,(2,1)} (s)=2 m \left( {{2 l_3}\over{m}} \right)^{2s}
\sin (\pi s)\, \Real \sum_{\nu=0}^{m/2-1}\int_0^\infty
{{dx}\over{1-e^{2 \pi x-2i \pi \nubar/m}}} \left[ \int_0^\infty
{{2 B dy}\over{(y^2+C^2)^s}} \right.$$ $$-\left. 2\int_0^\infty
{{dy}\over{(y^2+C^2)^s}} \left({{y-B}\over{e^{2 \pi (y-B)} e^{2
\pi i \nubar/m}-1}} - {{y+B}\over{e^{2 \pi (y+B)} e^{-2 \pi i
\nubar/m}-1}}\right)\right]\eql{zeta21s}$$

As for $\zeta_{e,1}$, the `anomalous' contribution of this term
vanishes. Moreover, $\zeta_{e,(2,1)}(1)=O(\l_3^2)$, again giving
no contribution to the effective action of the conformal field to
the order considered here.

Concerning the evaluation of the derivative at $s=0$, it is not
difficult to see that the first term in (\peq{zeta21s}) does not
contribute. So, we get
$$\zeta_{e,(2,1)}^\prime (0)=-4 \pi m\,
\Real \sum_{\nu=0}^{m/2-1} \int_0^\infty {{dx}\over{1-e^{2 \pi
x-2 \pi i \nubar/m}}} \int_0^\infty dy\,\,\times\hspace{*******}$$
$$ \hspace{************}\left({{y -
B}\over{e^{2 \pi (y-B)} e^{2 \pi i \nubar/m}-1}}-{{y+B}\over{e^{2
\pi (y+B)} e^{-2 \pi i \nubar/m}-1}}\right).\eql{zeta21p}$$

The $y$-integral leads to the polylogarithm function ${\rm
Li}_{\rm s}(x)=\sum_{n=1}^{\infty} x^n/n^s$,
$$\zeta_{e,(2,1)}^\prime (0)=-{{m}\over{\pi}}\, \Real
\sum_{\nu=0}^{m/2-1} \int_0^\infty {{dx}\over{1-e^{2 \pi x- 2 \pi
i \nubar/m}}} \left\{ {\rm Li}_2 (e^{-2 \pi i \nubar/m})+\right.
$$ $$\left. {\rm Li}_2 (e^{2 \pi i \nubar/m}) -2 {\rm Li}_2 (e^{2
\pi i \nubar/m - 2 \pi B }) - 4 \pi B  {\rm Li}_1 (e^{2 \pi i
\nubar/m-2 \pi B})+\pi^2 B^2 \right\}.\eql{zeta21poly}$$
\begin{ignore}
To simplify our notation, it is useful to define the projections
$$f_1=\sum_{\nu=0}^{m/2-1} \cos \left({{2l\nubar\pi}\over{m}}
\right) \cos \left({{2 k \nubar \pi}\over{m}}\right),\eql{f1}$$
$$f_2=\sum_{\nu=0}^{m/2-1} \sin \left({{2l\nubar\pi}\over{m}}
\right) \sin \left({{2 k \nubar \pi}\over{m}}\right),\eql{f2}$$
$$f_3=\sum_{\nu=0}^{m/2-1} \cos \left({{2k\nubar\pi}\over{m}}
\right).\eql{f3}$$

Performing the $x$-integral and using the previous
 definitions \mgn{Introduce BB and CB here} (\peq{f1}-\peq{f3}), a more 
compact expression for
$\zeta_{e,(2,1)}^\prime (0)$ is obtained $$\zeta_{e,(2,1)}^\prime
(0)={{m \bar{B}^2}\over{4 \pi^2}} \sum_{l=1}^{\infty}
\sum_{k=1}^{\infty} {{(f_1-f_2)}\over{k(\bar{B} l/2+k)^2}}
+{{m}\over{\pi^2}} \sum_{l=1}^{\infty} \sum_{k=1}^{\infty}
{{f_2}\over{k l}}+{{m \bar{B}^2}\over{8 \pi^2}}
\sum_{k=1}^{\infty} {{f_3}\over{k^3}}.\eql{zeta21x}$$

Following similar steps, we evaluate (\peq{zeta22}). Again the
result  $\zeta_{e,(2,2)} (0)=0$ is obtained and only higher
$l_3$-order contributions to $(\zeta_{e}^{\rm C})^\prime (0)$
appear \mgn{The notation for the conformal field has to be
introduced before}. Using definitions (\peq{f1}-\peq{f3}), the
contribution to the effective action reads
$$\zeta_{e,(2,2)}^\prime (0)=-2 \pi m \sum_{k=1}^\infty
\int_0^\infty dx {{\bar{C}^2 x^2}\over{4}} e^{-2 \pi k x} f_3 -$$
$$- 2 m \bar{B} \sum_{l=1}^\infty \sum_{k=1}^\infty
{{(f_1-f_2)}\over{l}} \int_0^\infty dx x e^{-\pi (2 k + \bar{B}
l)x} \left[\cos(\pi l \bar{C}x) -1\right]-$$
$$4m\sum_{l=1}^\infty \sum_{k=1}^\infty {{(f_1-f_2)}\over{l}}
\int_0^\infty dx {{\bar{C}x}\over2} e^{-\pi (2 k + \bar{B} l)x}
\sin(\pi l \bar{C}x)\eql{zeta22s}$$ $$-{{2 m}\over{\pi}}
\sum_{l=1}^\infty \sum_{k=1}^\infty {{(f_1-f_2)}\over{l^2}}
\int_0^\infty dx e^{-\pi (2 k + \bar{B} l)x} \left[\cos(\pi l
\bar{C}x) -1\right].$$

Expansion of the trigonometric functions will lead to the high
temperature expansion of the free energy we are interested in.
So, doing the $x$-integral in (\peq{zeta22s}) gives,
$$\zeta_{e,(2,2)}^\prime (0)=-{{m \bar{C}^2}\over{8 \pi^2}}
\sum_{k=1}^\infty {{f_3}\over{k^3}}-$$ $${{m}\over{\pi^2}}
\sum_{n=1}^\infty \sum_{l=1}^\infty \sum_{k=1}^\infty (-1)^n
\left( {{\bar{C}}\over2} \right)^{2n} l^{2n-2} (f_1-f_2)
{{\bar{B} l -k(2n-1)}\over{(\bar{B}
l/2+k)^{2+2n}}}.\eql{zeta22sx}$$

It is easy to show the identities, needed in equations
(\peq{zeta21x}) and (\peq{zeta22sx}), $$f_1-f_2={m\over2}
\sum_{t=1}^\infty (-1)^t \delta_{k+l,t m/2}, \eql{id1}$$
$$f_2=-{m\over4} \left( \sum_{t=1}^\infty (-1)^t \delta_{k+l,
mt/2} - \sum_{t=1}^\infty (-1)^t \delta_{k-l,-m t/2} -
\sum_{t=0}^\infty (-1)^t \delta_{k-l, m t/2} \right), \eql{id2}$$
$$f_3={m\over2} \sum_{t=1}^\infty (-1)^t \delta_{k, t
m/2},\eql{id3}$$ which allows one to obtain,
\end{ignore}
The $x$-integral can be performed by expanding the denominator in
exponentials and afterwards the $\nu$-summations can be done with the
help of
$$
\sum_{\nu = 0} ^ {m/2-1} e^{2\pi i \bar\nu n /m } =
{m \over 2} (-1)^k \delta_{n,km/2} , \, \,\,\,\, k\in\oZ .$$
After lengthy calculations one arrives at
 $$\eqalign{
 \zeta_{e,(2,1)}^\prime&
(0)=\bigg( - {{3(\bar{B}^2+4)}\over{8 m \pi^2}}
+{{m^2}\over{4\pi^2}} \bigg) \zeta_R (3)+{1\over6}\, \gamma\, \cr
&-{2\over{\pi^2}} \sum_{t=1}^\infty {{(-1)^t}\over{t^2}} \psi(m
t/2) + {{m^2}\over{\pi^2}} \sum_{t=1}^\infty (-1)^t \sum_{l=1}^{m
t/2-1} {{\bar{B}^2}\over{(t m - 2 l)(t m + (\bar{B} -2) l)^2}},\cr}
\eql{zeta21pr}$$
where $$\bar{B}=2-{{m^2}\over{8 \pi^2}} \beta^2$$
has been introduced. Proceeding similarly with $\zeta' _{e,(2,2)}
(0)$ we arrive at
$$\eqalign{
\zeta_{e,(2,2)}^\prime (0)&={{3
\bar{C}^2}\over{8m\pi^2}}\zeta_R (3)\cr
 &-{{m^2}\over{\pi^2}}
\sum_{n=1}^\infty (-1)^n \bar{C}^{2n} \sum_{t=1}^\infty (-1)^t
\sum_{l=1}^{t m/2-1} l^{2n-2} {{2 \bar{B} l-(t m - 2
l)(2n-1)}\over{((\bar{B}-2)l+m t)^{2n+2}}}\cr}\eql{zeta22pr}$$
with
$$\bar{C}^2=\left(4-{{m^2}\over{4 \pi^2}} \beta^2 \right)
{{m^2}\over{16 \pi^2}} \beta^2.$$
\begin{ignore}
Finally, expansion of expressions (\peq{zeta21pr}) and
(\peq{zeta22pr}) gives  the high temperature expansion
identifying $\beta=4\pi l_3/m$, and using, from (\peq{defs}),
$$\bar{B}=2-{{m^2}\over{8 \pi^2}} \beta^2, \qquad
\bar{C}^2=\left(4-{{m^2}\over{4 \pi^2}} \beta^2 \right)
{{m^2}\over{16 \pi^2}} \beta^2. \eql{deps}$$

\end{ignore}
Finally, expansion of expressions (\peq{zeta21pr}) and
(\peq{zeta22pr}) gives  the high temperature expansion
of both pieces of the second part
to the effective action, $$\zeta_{e,(2,1)}^\prime
(0)=\left(-{{3}\over{m \pi^2}}+{{m^2}\over{4 \pi^2}} \right)
\zeta_R (3) + \beta^2 \left( {{m^2}\over{48 \pi^2}} -
{{3m}\over{16 \pi^4}} \zeta_R (3) \right) +
O(\beta^4)\eql{zeta21final}$$ and
$$\eqalign{\zeta_{e,(2,2)}^\prime (0)= \beta^2 \left(-
{{m^2}\over{192 \pi^2}} + {{3m}\over{16 \pi^4}} \zeta_R (3)
\right) + O(\beta^4)}.\eql{zeta22final}$$

To complete our calculation, the oblate-only contribution,
(\peq{i3}), to the effective action must be evaluated. It is
straightforward to see that a change of variables gives
 $$\zeta_{e,3} (s)={{2^{2s+1}}\over{m}} i \tan \theta
\sum_{\nu=0}^{m/2-1} \sum_{p=0}^{\infty} n^{2-2s} \int_{\rm C}
{{1}\over{1-e^{2 \pi \zeta \tan \theta n/m}}}
{{d\zeta}\over{(1-\zeta^2)^s}}\eql{zeta3}$$
where the label $n$, see (2.7), is reintroduced,
and where $C$ is a clockwise contour enclosing the cut at
$\xi \in [1,\infty)$ in the complex $\xi$-plane.

Again $\zeta_{e,3} (0) = 0$, but now, the conformally required
 term $\zeta_{e,3}(1)$
{\it is} relevant to the evaluated order $\beta^2$. We will return
to its calculation after the analysis of the derivative.

Writing (\peq{zeta3}) as a real integral and introducing
$$\beta^\prime = {{4 \pi \tan \theta}\over{m}} = \beta \left(
1-{{m ^2}\over{16 \pi^2}} \beta^2 \right)^{-1/2}\eql{betap}$$ we
get, $$\zeta_{e,3}^\prime (0) = -\beta^\prime \sum_{\nu=0}^{m/2-1}
\sum_{p=0}^\infty n^2 \int_1^\infty {{d\eta}\over{1-e^{\eta
\beta^\prime n/2}}}.\eql{zeta3p}$$

The denominator can be given as a geometric
series and considering the integral
representation for the exponential, $$e^{-x} = {1\over{2 \pi i}}
\int_{c-i \infty}^{c+i \infty} x^{-s} \Gamma (s) ds,
\qquad \Real c > 0,\eql{expo}
 $$
we immediately obtain the expression $$\zeta_{e,3}^\prime (0) = 4
\sum_{\rm poles} {\rm Res} \left[ (\beta^\prime)^{-s} \zeta_H
(s-1,1/2) \Gamma (s) \zeta_R (s+1)\right]\eql{zeta3e}$$ where
terms of different order in $\beta$ are related to successive
poles in the expression in square brackets.

Then, the final result for the third contribution is
$$\zeta_{e,3}^\prime (0)= {4\over{\beta^2}} \zeta_R (3) -
{{m^2}\over{4 \pi^2}} \zeta_R (3) - {1\over6} \log 2 -{1\over6}
\log \beta - 2 \zeta_R^\prime (-1)$$
$$\hspace{***********}-{{m^2 \beta^2}\over{192
\pi^2}} + {{7 \beta^2}\over{5760}}+O(\beta^4).\eql{zeta3final}$$

A similar procedure allows one to calculate the extra
contribution needed for conformal fields previously mentioned,
 $$\eqalign{
 \zeta_{e,3} (1) &= 2
\beta^\prime \sum_{\rm poles} {\rm Res} \big[
(\beta^\prime)^{-s} \Gamma (s) \zeta_H (s,1/2) \zeta_R (s)
\big]\cr
&=2 \left( \gamma + 2 \log 2 - \log \beta \right) +
O(\beta^2).\cr}\eql{zeta31final}$$

Putting all the results (\peq{z1p}), (\peq{zeta21final}),
(\peq{zeta22final}), (\peq{zeta3final}) and (\peq{zeta31final})
together, our complete final expressions for the $m$-even case
turn out to be,
$$\zeta_e^\prime (0) = {4\over{\beta^2}} \zeta_R
(3) - {1\over6} \log 2 - {1\over6} \log \beta - 2 \zeta_R^\prime
(-1) + \beta^2 \left( {{m^2}\over{96 \pi^2}} + {7\over{5760}}
\right) + O(\beta^4),\eql{final}$$
and for the conformal case,
$$\eqalign{
{\zeta_e^{\rm C}}' (0) &= {4\over{\beta^2}} \zeta_R (3) -
{1\over6} \log 2 - {1\over6} \log \beta - 2 \zeta_R^\prime (-1)
\cr
&+\,\beta^2 \left( {{m^2}\over{96 \pi^2}} + {7\over{5760}} +
{{m^2}\over{128 \pi^2}} \left( \gamma + 2 \log 2 - \log \beta
\right) \right) + O(\beta^4).\cr}\eql{finalconf}$$

In giving these calculational details we have provided the
necessary minimum for anyone who wishes to check our results or
who might want to extend them.

\section{\bf Results. Thermal interpretation.}

We now collect the expressions for the high temperature series of
the free energies, $\Phi(\be)$, of all the fields considered in
this paper and these are our main results.
They are valid for all $m$, even and odd.

For the scalar field, $$
\Phi(\be)\sim
-{2\over\be^3}\ze_R(3)+{1\over12\be}\log\be+{1\over\be}\ze_R'(-1)-{1\over2}
\bigg({m^2\over96\pi^2}+{7\over5760}\bigg)\be+O(\be^3).
\eql{frensc}$$

For the conformal (in three dimensions) scalar field,
$$\eqalign{
\Phi(\be)&\sim
-{2\over\be^3}\ze_R(3)+{1\over12\be}\log\be+{1\over\be}\ze_R'(-1)\cr
&\hspace{*****}-{1\over2}\bigg({m^2\over96\pi^2}+{7\over5760}+
{m^2\over128\pi^2}(\ga+2\log2 -\log\be)\bigg)\be+O(\be^3).\cr}
\eql{frencsc}$$

For the periodic spinor field, $$\eqalign{
\Phi(\be)&\sim{4\over\be^3}\ze_R(3)+{1\over3\be}\log\be+{4\over\be}\ze_R'(-1)+
\bigg({m^2\over96\pi^2}-{1\over720}\bigg)\be\cr &\hspace{****}+
{m^2\be\over128\pi^2}(\ga-\log\be)+O(\be^2).\cr} \eql{frensp}$$

For {\it twisted} (\ie antiperiodic) spinors
$$
\Phi(\be)\sim-{3\ze_R(3)\over\be^3}+{1\over3\be}\log2-{\be\over240}
-{m^2\be\over
128\pi^2}\log2+O(\be^2), \eql{frentwsp}$$ which, as anticipated,
has a legitimate thermodynamic sign. Note the absence of
$\log\be/\be$ and $\ze_R'(-1)/\be$ terms. The former means that there is
no term proportional to $T$ is the high--T expansion of the internal
energy, which is typical of standard spinor thermodynamics in two dimensions.

The unidentified case is obtained by setting $m$ equal to unity.

\section{\bf Discussion of results}
Our first comment is to remark on the sign of the leading term in
the periodic  spinor expression, (\peq{frensp}). This differs from that in
[\pref{Dow2}] which was the result of a basic error. The twisted
case, (\peq{frentwsp}), exhibits a `thermal sign'.

The scalar result, (\peq{frensc}), also differs from that in [\pref{Dow2}];
again a  consequence of sign errors in that article.

For comparison, the free energy resulting from the small $l_3$
expansion of the action of the AdS-Taub Bolt instanton, rendered
finite by subtraction of the AdS-Taub Nut value, is given by
[\pref{HHP}], $$
\Phi\sim-\al\bigg({1\over\be^3}-{9\over8\pi^2\be}
+{27(m+2)^2\over1024\pi^4}\be\bigg)
\eql{holo}$$ where $\al$ is a constant.

The precise connection between (\peq{holo}) and the quantities
evaluated in this paper is problematic. In particular, while the
leading term is comparable, there are no transcendental
quantities and the dependence on the modding integer, $m$, is
different. Nor does there seem to be any hope of patching up
these differences by taking combinations of fields. Of course
expression (\peq{holo}) is subject to the uncertainties of the
regularising  subtraction.

\section{\bf Conclusion}

For conformal scalars and spinors, the awkward form of
the eigenvalues seems to necessitate an expansion. This is of no
real concern if all we want is the asymptotic behaviour in the
extreme oblate case but does prevent a computation of the
effective action for {\it any} value of the squashing parameter,
which is something one ought to be able to find. In the scalar
field case, for a particular choice of propagation operator, it
is possible to calculate the effective action for the complete
range of $l_3$.

In the course of the calculation several summations arise that
have been encountered in other connections. Some extensions of
these earlier results are necessary.

The details and manipulations of the spinor calculation may have
application in string theory where the calculation of the
spectral asymmetry invariant, $\eta(0)$, for twisted spinors on a
space whose boundary is a squashed lens space appears
[\pref{GandH}]. In this case one would also like to evaluate the
determinants for the more general periodicity condition
(\peq{pperiod}), (\peq{twrestr}).

The entire calculation could be repeated for the arbitrary odd-dimensional
sphere by making use of the results of B\"ar, [\pref{Bar1,Bar2}],
and their extension. In particular it would be relatively straightforward
to evaluate the poles of the \zf\ and hence the \hk\ expansion coefficients
in terms of generalised Bernoulli functions. The corresponding questions
for the spectral asymmetry function, $\eta(s)$, can also be answered; \cf
[\pref{Dow1}] for the three-dimensional, unidentified case.
\vskip5truept
\noindent{\bf References}
\vskip 5truept
\begin{putreferences}
\ref{APS}{Atiyah,M.F., V.K.Patodi and I.M.Singer: Spectral
asymmetry and Riemannian geometry \mpcps{77}{75}{43}.}
\ref{AandD}{Apps,J.S. and Dowker,J.S. \cqg{15}{98}{1121}.}
\ref{AandT}{Awada,M.A. and D.J.Toms: Induced gravitational and
gauge-field actions from quantised matter fields in non-abelian
Kaluza-Klein thory \np{245}{84}{161}.} \ref{BandI}{Baacke,J. and
Y.Igarishi: Casimir energy of confined massive quarks
\prD{27}{83}{460}.} \ref{Barnesa}{Barnes,E.W.: On the Theory of
the multiple Gamma function {\it Trans. Camb. Phil. Soc.} {\bf
19} (1903) 374.} \ref{Barnesb}{Barnes,E.W.: On the asymptotic
expansion of integral functions of multiple linear sequence, {\it
Trans. Camb. Phil. Soc.} {\bf 19} (1903) 426.}
\ref{Bar1}{B\"ar,C. {\it Arch.d.Math.}{\bf 59} (1992) 65.}
\ref{Bar2}{B\"ar,C. {\it Geom. and Func. Anal.} {\bf 6} (1996) 899.}
\ref{Barv}{Barvinsky,A.O. Yu.A.Kamenshchik and I.P.Karmazin:
One-loop quantum cosmology \aop {219}{92}{201}.}
\ref{BandM}{Beers,B.L. and Millman, R.S. :The spectra of the
Laplace-Beltrami operator on compact, semisimple Lie groups.
\ajm{99}{1975}{801-807}.} \ref{BandH}{Bender,C.M. and P.Hays:
Zero point energy of fields in a confined volume
\prD{14}{76}{2622}.} \ref{BBG}{Bla\v zi\' c,N., Bokan,N. and
Gilkey,P.B.: Spectral geometry of the form valued Laplacian for
manifolds with boundary \ijpam{23}{92}{103-120}}
\ref{BEK}{Bordag,M., E.Elizalde and K.Kirsten: { Heat kernel
coefficients of the Laplace operator on the D-dimensional ball},
\jmp{37}{96}{895}.} \ref{BGKE}{Bordag,M., B.Geyer, K.Kirsten and
E.Elizalde,: { Zeta function determinant of the Laplace operator
on the D-dimensional ball}, \cmp{179}{96}{215}.}
\ref{BKD}{Bordag,M., K.Kirsten,K. and Dowker,J.S.: Heat kernels
and functional determinants on the generalized cone
\cmp{182}{96}{371}.} \ref{Branson}{Branson,T.P.: Conformally
covariant equations on differential forms \cpde{7}{82}{393-431}.}
\ref{BandG2}{Branson,T.P. and P.B.Gilkey {\it Comm. Partial Diff.
Eqns.} {\bf 15} (1990) 245.} \ref{BGV}{Branson,T.P., P.B.Gilkey
and D.V.Vassilevich {\it The Asymptotics of the Laplacian on a
manifold with boundary} II, hep-th/9504029.}
\ref{BCZ1}{Bytsenko,A.A, Cognola,G. and Zerbini, S. : Quantum
fields in hyperbolic space-times with finite spatial volume,
hep-th/9605209.} \ref{BCZ2}{Bytsenko,A.A, Cognola,G. and Zerbini,
S. : Determinant of Laplacian on a non-compact 3-dimensional
hyperbolic manifold with finite volume, hep-th /9608089.}
\ref{CandH2}{Camporesi,R. and Higuchi, A.: Plancherel measure for
$p$-forms in real hyperbolic space, \jgp{15}{94}{57-94}.}
\ref{CandH}{Camporesi,R. and A.Higuchi {\it On the eigenfunctions
of the Dirac operator on spheres and real hyperbolic spaces},
gr-qc/9505009.} \ref{ChandD}{Chang, Peter and J.S.Dowker :Vacuum
energy on orbifold factors of spheres, \np{395}{93}{407}.}
\ref{cheeg1}{Cheeger, J.: Spectral Geometry of Singular
Riemannian Spaces. \jdg {18}{83}{575}.} \ref{cheeg2}{Cheeger,J.:
Hodge theory of complex cones {\it Ast\'{e}risque} {\bf
101-102}(1983) 118-134} \ref{Chou}{Chou,A.W.: The Dirac operator
on spaces with conical singularities and positive scalar
curvature, \tams{289}{85}{1-40}.} \ref{CandT}{Copeland,E. and
Toms,D.J.: Quantized antisymmetric tensor fields and
self-consistent dimensional reduction in higher-dimensional
spacetimes, \break\np{255}{85}{201}} \ref{DandH}{D'Eath,P.D. and
J.J.Halliwell: Fermions in quantum cosmology \prD{35}{87}{1100}.}
\ref{cheeg3}{Cheeger,J.:Analytic torsion and the heat equation.
\aom{109} {79}{259-322}.} \ref{CEK}{Cognola,Elizalde and
Kirsten,K} \ref{DandE}{D'Eath,P.D. and G.V.M.Esposito: Local
boundary conditions for Dirac operator and one-loop quantum
cosmology \prD{43}{91}{3234}.} \ref{DandE2}{D'Eath,P.D. and
G.V.M.Esposito: Spectral boundary conditions in one-loop quantum
cosmology \prD{44}{91}{1713}.} \ref{Dow3}{Dowker,J.S.: Effective
action on spherical domains, \cmp{162}{94} {633}.}
\ref{Dow1}{Dowker,J.S. `Vacuum energy in a squashed Einstein
Universe' in {\it Quantum Gravity}, ed. S.C.Christensen (I.O.P.,
1984).} \ref{Dow2}{Dowker,J.S. \cqg{16}{99}{1937}.}
\ref{Dow8}{Dowker,J.S.{\it Robin conditions on the Euclidean
ball} MUTP/95/7; hep-th\break/9506042. {\it Class. Quant.Grav.}
to be published.} \ref{Dow9}{Dowker,J.S. {\it Oddball
determinants} MUTP/95/12; hep-th/9507096.}
\ref{Dow10}{Dowker,J.S. {\it Spin on the 4-ball}, hep-th/9508082,
{\it Phys. Lett. B}, to be published.} \ref{DandA2}{Dowker,J.S.
and J.S.Apps, {\it Functional determinants on certain domains}.
To appear in the Proceedings of the 6th Moscow Quantum Gravity
Seminar held in Moscow, June 1995; hep-th/9506204.}
\ref{DandK}{Dowker,J.S. and Kennedy,G. \jpa{11}{78}{895}.}
\ref{DandB}{Dowker,J.S. and Banach,R. \jpa{11}{78}{2255}.}
\ref{DABK}{Dowker,J.S., Apps,J.S., Bordag,M. and Kirsten,K.:
Spectral invariants for the Dirac equation with various boundary
conditions {\it Class. Quant.Grav.} to be published,
hep-th/9511060.} \ref{EandR}{E.Elizalde and A.Romeo : An integral
involving the generalized zeta function, {\it International J. of
Math. and Phys.} {\bf13} (1994) 453.} \ref{ELV2}{Elizalde, E.,
Lygren, M. and Vassilevich, D.V. : Zeta function for the laplace
operator acting on forms in a ball with gauge boundary
conditions. hep-th/9605026} \ref{ELV1}{Elizalde, E., Lygren, M.
and Vassilevich, D.V. : Antisymmetric tensor fields on spheres:
functional determinants and non-local counterterms, \jmp{}{96}{}
to appear. hep-th/ 9602113} \ref{Kam2}{Esposito,G.,
A.Y.Kamenshchik, I.V.Mishakov and G.Pollifrone: Gravitons in
one-loop quantum cosmology \prD{50}{94}{6329};
\prD{52}{95}{3457}.}
\ref{Erdelyi}{A.Erdelyi,W.Magnus,F.Oberhettinger and F.G.Tricomi
{\it Higher Transcendental Functions} Vol.I McGraw-Hill, New
York, 1953.} \ref{Esposito}{Esposito,G.: { Quantum Gravity,
Quantum Cosmology and Lorentzian Geometries}, Lecture Notes in
Physics, Monographs, Vol. m12, Springer-Verlag, Berlin 1994.}
\ref{Esposito2}{Esposito,G. {\it Nonlocal properties in Euclidean
Quantum Gravity}. To appear in Proceedings of 3rd. Workshop on
Quantum Field Theory under the Influence of External Conditions,
Leipzig, September 1995; gr-qc/9508056.} \ref{EKMP}{Esposito G,
Kamenshchik Yu A, Mishakov I V and Pollifrone G.: One-loop
Amplitudes in Euclidean quantum gravity. \prD {52}{96}{3457}.}
\ref{ETP}{Esposito,G., H.A.Morales-T\'{e}cotl and L.O.Pimentel {\it
Essential self-adjointness in one-loop quantum cosmology},
gr-qc/9510020.} \ref{FORW}{Forgacs,P., L.O'Raifeartaigh and
A.Wipf: Scattering theory, U(1) anomaly and index theorems for
compact and non-compact manifolds \np{293}{87}{559}.}
\ref{GandM}{Gallot S. and Meyer,D. : Op\'{e}rateur de coubure et
Laplacian des formes diff\'{e}ren-\break tielles d'une vari\'{e}t\'{e}
riemannienne \jmpa{54}{1975}{289}.} \ref{GandH}{Gauntlett,J.P.
and Harvey,J.A. \np{463}{96}{287}.}
\ref{Gibb}{Gibbons,G.W.\pl{60A}{77}{385}.} \ref{GPR}{Gibbons,
G.W, Pope, C, and R\"omer,H, \np{157}{79}{377}.}
\ref{Gilkey1}{Gilkey, P.B, Invariance theory, the heat equation
and the Atiyah-Singer index theorem, 2nd. Edn., CTC Press, Boca
Raton 1995.} \ref{Gilkey2}{Gilkey,P.B.:On the index of geometric
operators for Riemannian manifolds with boundary
\aim{102}{93}{129}.} \ref{Gilkey3}{Gilkey,P.B.: The boundary
integrand in the formula for the signature and Euler
characteristic of a manifold with boundary \aim{15}{75}{334}.}
\ref{Grubb}{Grubb,G. {\it Comm. Partial Diff. Eqns.} {\bf 17}
(1992) 2031.} \ref{GandS1}{Grubb,G. and R.T.Seeley
\cras{317}{1993}{1124}; \invm{121}{95} {481}.}
\ref{GandS}{G\"unther,P. and Schimming,R.:Curvature and spectrum
of compact Riemannian manifolds, \jdg{12}{77}{599-618}.}
\ref{HHP}{Hawking,S.W., Hunter,C.J. and Page,D.N.
\prD{59}{99}{044033}.} \ref{IandT}{Ikeda,A. and
Taniguchi,Y.:Spectra and eigenforms of the Laplacian on $S^n$ and
$P^n(C)$. \ojm{15}{1978}{515-546}.} \ref{IandK}{Iwasaki,I. and
Katase,K. :On the spectra of Laplace operator on $\La^*(S^n)$
\pja{55}{79}{141}.} \ref{JandK}{Jaroszewicz,T. and P.S.Kurzepa:
Polyakov spin factors and Laplacians on homogeneous spaces
\aop{213}{92}{135}.} \ref{Kam}{Kamenshchik,Yu.A. and
I.V.Mishakov: Fermions in one-loop quantum cosmology
\prD{47}{93}{1380}.} \ref{KandM}{Kamenshchik,Yu.A. and
I.V.Mishakov: Zeta function technique for quantum cosmology {\it
Int. J. Mod. Phys.} {\bf A7} (1992) 3265.} \ref{KandC}{Kirsten,K.
and Cognola.G,: { Heat-kernel coefficients and functional
determinants for higher spin fields on the ball} \cqg{13}{96}
{633-644}.} \ref{Levitin}{Levitin,M.: { Dirichlet and Neumann
invariants for Euclidean balls}, {\it Diff. Geom. and its Appl.},
to be published.} \ref{Luck}{Luckock,H.C.: Mixed boundary
conditions in quantum field theory \jmp{32}{91}{1755}.}
\ref{MandL}{Luckock,H.C. and Moss,I.G,: The quantum geometry of
random surfaces and spinning strings \cqg{6}{89}{1993}.}
\ref{Ma}{Ma,Z.Q.: Axial anomaly and index theorem for a
two-dimensional disc with boundary \jpa{19}{86}{L317}.}
\ref{Mcav}{McAvity,D.M.: Heat-kernel asymptotics for mixed
boundary conditions \cqg{9}{92}{1983}.}
\ref{MandV}{Marachevsky,V.N. and D.V.Vassilevich {\it
Diffeomorphism invariant eigenvalue \break problem for metric
perturbations in a bounded region}, SPbU-IP-95, \break
gr-qc/9509051.} \ref{Milton}{Milton,K.A.: Zero point energy of
confined fermions \prD{22}{80}{1444}.}
\ref{MandS}{Mishchenko,A.V. and Yu.A.Sitenko: Spectral boundary
conditions and index theorem for two-dimensional manifolds with
boundary \aop{218}{92}{199}.} \ref{Moss}{Moss,I.G.
\cqg{6}{89}{759}.} \ref{MandP}{Moss,I.G. and S.J.Poletti:
Conformal anomaly on an Einstein space with boundary
\pl{B333}{94}{326}.} \ref{MandP2}{Moss,I.G. and S.J.Poletti
\np{341}{90}{155}.} \ref{NLS}{Nesterenko,V.V., Lambiase,G. and
Scarpetta, hep-th/00.} \ref{NandOC}{Nash, C. and O'Connor,D.J.:
Determinants of Laplacians, the Ray-Singer torsion on lens spaces
and the Riemann zeta function \jmp{36}{95}{1462}.}
\ref{NandS}{Niemi,A.J. and G.W.Semenoff: Index theorem on open
infinite manifolds \np {269}{86}{131}.} \ref{NandT}{Ninomiya,M.
and C.I.Tan: Axial anomaly and index thorem for manifolds with
boundary \np{245}{85}{199}.} \ref{norlund2}{N\"orlund~N.
E.:M\'{e}moire sur les polynomes de Bernoulli. \am {4}{21} {1922}.}
\ref{Poletti}{Poletti,S.J. \pl{B249}{90}{355}.}
\ref{Pope}{Pope,C.\jpa{14}{81}{L133}.} \ref{RandT}{Russell,I.H.
and Toms D.J.: Vacuum energy for massive forms in $R^m\times
S^N$, \cqg{4}{86}{1357}.} \ref{RandS}{R\"omer,H. and P.B.Schroer
\pl{21}{77}{182}.} \ref{Trautman}{Trautman,A.: Spinors and Dirac
operators on hypersurfaces \jmp{33}{92}{4011}.}
\ref{Unwin}{Unwin, S.D., PhD thesis, University of Manchester
1980.} \ref{Vass}{Vassilevich,D.V.{Vector fields on a disk with
mixed boundary conditions} gr-qc /9404052.} \ref{Voros}{Voros,A.:
Spectral functions, special functions and the Selberg zeta
function. \cmp{110}{87}439.} \ref{Ray}{Ray,D.B.: Reidemeister
torsion and the Laplacian on lens spaces \aim{4}{70}{109}.}
\ref{McandO}{McAvity,D.M. and Osborne,H. Asymptotic expansion of
the heat kernel for generalised boundary conditions
\cqg{8}{91}{1445}.} \ref{AandE}{Avramidi,I. and Esposito,G.}
\ref{MandS}{Moss,I.G. and Silva P.J. Invariant boundary
conditions for gauge theories gr-qc/9610023.}
\ref{barv}{Barvinsky,A.O.\pl{195B}{87}{344}.}
\end{putreferences}

\bye